# Centroid Molecular Dynamics Can Be Greatly Accelerated Through Neural Network Learned Centroid Forces Derived from Path Integral Molecular Dynamics

Timothy D. Loose, Patrick G. Sahrmann, and Gregory A. Voth*

Department of Chemistry, Chicago Center for Theoretical Chemistry, James Franck Institute, and Institute for Biophysical Dynamics, The University of Chicago, Chicago, IL 60637, USA

*Corresponding Author: gavoth@uchicago.edu

**Abstract**

For nearly the past 30 years, Centroid Molecular Dynamics (CMD) has proven to be a viable classical-like phase space formulation for the calculation of quantum dynamical properties. However, calculation of the centroid effective force remains a significant computational cost and limits the ability of CMD to be an efficient approach to study condensed phase quantum dynamics. In this paper we introduce a neural network-based methodology for first learning the centroid effective force from path integral molecular dynamics data, which is subsequently used as an effective force field to evolve the centroids directly with the CMD algorithm. This method, called Machine-Learned Centroid Molecular Dynamics (ML-CMD) is faster and far less costly than both standard "on the fly" CMD and ring polymer molecular dynamics (RPMD). The training aspect of ML-CMD is also straightforwardly implemented utilizing the DeepMD software kit. ML-CMD is then applied to two model systems to illustrate the approach: liquid para-hydrogen and water. The results show comparable accuracy to both CMD and RPMD in the estimation of quantum dynamical properties, including the self-diffusion constant and velocity time correlation function, but for significantly reduced overall computational cost.

**I. Introduction**

The accurate simulation of quantum dynamics is limited by the computational complexity of solving the time-dependent Schrödinger equation. While classical molecular dynamics (MD) based on empirical force fields are capable of utilizing certain information from quantum calculations, the treatment of the nuclei as point particles as well as the inherent limitations of a pairwise decomposable description of the intermolecular interactions limits their accuracy.[1,2] The most straightforward approach to address this accuracy problem is to perform full electronic structure calculations for each timestep to calculate the force vectors on each particle, such as in the case of *ab initio* molecular dynamics (AIMD).[3,4] However such methods can be prohibitively expensive for all but the smallest systems. Additionally, these calculations must be performed repeatedly over the course of a simulation which highly limits the timescales accessible to study with these methods. Many alternatives have been employed to limit the cost of representing the effects of quantum electronic structure within MD simulations, such as treating the majority of the system with classical mechanics and saving the quantum mechanical calculations for specific regions as in hybrid quantum mechanics/molecular mechanics.[5]

Irrespective of the underlying accuracy of the representation of the forces on the system nuclei, the challenge of quantum dynamics is compounded by the fact that for many systems of interest exhibit nuclear quantum effects (NQEs), even at thermal equilibrium. Path Integral (PI) methods can capture NQEs in such equilibrium circumstances. This technique relies on the imaginary time formulation of Feynman path integral quantum mechanics, in which the classical principle of least action is generalized to quantum systems via functional integration over all paths that system can take between an initial and final point.[6] In the case of molecular simulation, PI based methods re-cast a quantum mechanical description of the system of interest into an isomorphic classical one where each quantum particle is represented by a set of $P$ classical quasiparticles or "beads", for which standard MD algorithms can be utilized.[7,8] In particular, it can be shown that, in the P → ∞ limit, the static equilibrium properties of a quantum particle can be described by the Boltzmann statistics generated by the following Hamiltonian:

$$H = \sum_{n=1}^{P}\left[\frac{p_n^2}{2m_n'} + \frac{1}{2}m\omega_P^2(x_n - x_{n+1})^2 + \frac{U(x_n)}{P}\right] \quad (1)$$

where $m_n'$ are fictitious mass parameters, $\omega_P^2 = \sqrt{P}/\beta\hbar$, and $P$ is the total number of replicas or beads chosen for the discretized imaginary time path. Each bead represents the particle at a discrete position in imaginary time, so $x_{P+1} = x_1$ in order to guarantee that only paths which begin and end in the same place are considered for each configuration. This Hamiltonian effectively describes a collection of classical-like particles each acting under a potential $U(x)/P$ that are attached by harmonic oscillators to the adjacent particles in a "ring polymer" or "necklace". (Note that the notation here is for a single quantized particle in a one-dimensional potential, but the notation is readily generalized to more than one particle in three dimensions.) Molecular dynamics simulations performed using the Hamiltonian in equation 1 is Path Integral Molecular Dynamics (PIMD). PIMD provides a route to calculating quantum static equilibrium properties in a computationally feasible fashion, generally achieving converged results at P~30 replicas for a system such as liquid water at ambient temperature.[8] However, the PIMD Hamiltonian cannot be used to estimate quantum dynamics and must be regarded only as a sampling tool for equilibrium statistics.[7,8] The PIMD approach is also still significantly slower than classical MD due to the increased complexity of the system being simulated, but largely scales the same way as does classical MD in terms of computational cost (i.e., there is simply a "cost" prefactor proportional to the value of $P$).

Centroid Molecular Dynamics (CMD), introduced by Cao and Voth nearly thirty years ago, provides a means to estimate certain quantum dynamical information from the discretized imaginary time path integral via the dynamics of the imaginary time path centroid moving in a classical-like fashion under the mean centroid force.[9,10] This method is motivated by Feynman's observation that the imaginary time path centroid is the most classical-like variable of a quantum system.[6,11] The original papers introducing the CMD concept replied largely on *ad hoc* arguments to justify the method, but two subsequent papers[12,13] in 1999 provided an exact formulation of centroid quantum dynamics and also a route to deriving CMD as an approximation to those exact dynamics. These latter two papers are sometimes not cited by authors when discussing CMD so the primary content of that work is briefly reviewed here for completeness.

Formally, a quasi-density operator (QDO) for the centroid density can be defined which can be used to formulate the exact dynamics of the imaginary time path centroids.[12] In one dimension, this QDO is equal to

$$\varphi(x_c, p_c) = \frac{\hbar}{2\pi} \int_{-\infty}^{\infty} d\zeta \int_{-\infty}^{\infty} d\eta \, e^{i\zeta(\hat{x}-x_c)+i\eta(\hat{p}-p_c)-\beta\hat{H}}, \qquad (2)$$

where $x_c$ and $p_c$ are centroid positions and momenta, $\hat{H}$ is the system's hamiltonian and $\beta = 1/k_B T$. The centroid distribution function can be obtained by tracing this operator

$$\rho(x_c, p_c) = Tr[\varphi(x_c, p_c)]. \tag{3}$$

Evaluation of this trace gives a classical-like form for the centroid distribution function which separates the position and momentum components when a Cartesian coordinate system is used, such that

$$\rho(x_c, p_c) = e^{-\frac{\beta p_c^2}{2m}} \rho(x_c) = e^{-\frac{\beta p_c^2}{2m}} e^{-\beta V_c(x_c)} \tag{4}$$

where $V_c(x_c)$ is the effective centroid quantum potential of mean force. Integrating over the centroid position and momentum variables thus yields the standard quantum partition function

$$Z = \int \int \frac{dx_c dp_c}{2\pi\hbar} \rho(x_c, p_c). \tag{5}$$

This partition function can be used to calculate the average of a physical observable corresponding to an operator $\hat{A}$ as

$$\langle \hat{A} \rangle = \frac{1}{Z} \int \int \frac{dx_c dp_c}{2\pi\hbar} \rho(x_c, p_c) \hat{A}(x_c, p_c; t), \tag{6}$$

where the time dependent centroid variable $A_c(x_c, p_c; t)$ is defined as

$$\hat{A}(x_c, p_c; t) = Tr\left[\hat{\varphi}(x_c, p_c) e^{\frac{i\hat{H}t}{\hbar}} \hat{A} e^{-\frac{i\hat{H}t}{\hbar}}\right] / \rho(x_c, p_c). \tag{7}$$

For the exact dynamics of the centroid variables, a normalized time dependent QDO can also be defined as

$$\hat{\delta}_c(t; x_c, p_c) = e^{-\frac{i\hat{H}'t}{\hbar}} \hat{\delta}_c(x_c, p_c) e^{\frac{i\hat{H}'t}{\hbar}} \qquad (8)$$

$$\hat{\delta}_c(x_c, p_c) = \hat{\varphi}(x_c, p_c)/\rho(x_c, p_c). \qquad (9)$$

where $\hat{H}'$ is a time-independent Hamiltonian upon which the system evolves ($\hat{H}' = \hat{H}$ in the usual equilibrium case). It is important to note that the QDO in this derivation is not a typical density operator: while it is Hermitian with non-negative diagonal elements, it is not positive-definite. This operator can be used to describe the exact dynamics of the path centroids; however this is not a useful approach for non-trivial systems for which the quantum Liouville equation cannot be solved.[12] Instead, various approximations to the QDO can be made which result in various forms of path integral based methods including linearized quantum dynamics, centroid Hamiltonian dynamics, and CMD.[13] In the case of CMD the centroid phase space variables are propagated quasi-classically by virtue of the following *ansatz* :

$$\hat{\delta}_c(t; x_c, p_c) \approx \hat{\delta}_c(x_c(t), p_c(t)) \qquad (10)$$

$$m\dot{x}_c(t) = p_c(t); \; \dot{p}_c(t) \approx F_c(x_c(t)) = F_{CMD}(t) \qquad (11)$$

The approximation made here assumes that the QDO is the same at $t = 0$ as at later times except for the placement of the centroids. This mean field-like assumption is reasonable for cases in which linear response theory approximates the dynamics of the system well, and for systems that have strong regression to equilbrium behavior. This perspective reveals that CMD is likely to be most accurate for systems at equilibrium and in which coherent (purely quantum) aspects of the dynamics are not likely to have enough time to influence the system significantly before the correlations die out (de-cohere)

The key to CMD is thus to numerically calculate the effective equilibrium forces felt by the path centroids in one way or another. Sampling of the non-centroid imaginary time path integral modes at each centroid position determines an effective potential on which to propagate the centroids in a classical-like fashion. However, fully sampling these modes at each timestep in a simulation is usually very computationally expensive and so it is arguably this feature of CMD that has precluded its application to certain problems over the years. Numerical implementations of CMD instead attempt to adiabatically separate the centroid and non-centroid imaginary time path integral (Matsubara) modes, of which the zero frequency mode is the centroid. Adiabatic separation is achieved by setting the fictitious masses of the non-centroid modes to be much lower than that of the centroid, which is set to the physical mass of the particle. Then, one attaches thermostats to the non-centroid modes to help more rapidly sample them. While the adiabatic approximation enables "on-the-fly" calculation of the centroid effective force, generating centroid trajectories in CMD still involves significant computational overhead.

As an alternative to CMD, Manolopoulos and co-workers subsequently introduced ring polymer molecular dynamics (RPMD) as another approximate quantum dynamics approach,[14–16] which shared certain key aspects of the spirit of CMD. In RPMD, the fictitious masses of the ring polymer beads are set to the physical mass of the particle, and each bead is evolved as a dynamical variable with PIMD. One then makes the *ad hoc* argument that the MD sampling time in PIMD (as measured by integrator timesteps) is related to the actual real time of the quantum dynamics.

The RPMD approach thus removes the requirement that any sort of centroid force averaging be carried out as in CMD. It should be noted that a subsequent analysis[17] showed that RPMD has no clear connection to real time quantum dynamics, but the methodology remains popular among users given its ease of use. In addition, an analysis of an approximate, but to date impractical form of quantum dynamics called "Matsubara Dynamics" has suggested that RPMD and CMD can be related to that approximation through further approximations.[18] Also of note is that RPMD is numerically faster than CMD, but the increase in speed is relatively small in comparison to classical MD, which is significantly faster than both.

While both RPMD and CMD can capture a range of quantum effects such as incoherent tunneling and zero-point quantization, they are not without their drawbacks. For example for certain potential energy functions and at low enough temperatures, CMD may exhibit a "curvature problem", which has a tendency to red shift certain vibrational frequencies for some systems,[19] This behavior can make CMD less suitable for spectrum prediction for certain systems, although this issue generally vanishes at room temperature or higher.[20] On the other hand, RPMD suffers from a spurious resonance problem, in which the centroid dynamics becomes coupled to the harmonic oscillations of the ring polymer, introducing artificial resonances into the spectra.[21] Thermostatted RPMD (TRPMD), in which Langevin thermostats are attached to the ring polymer internal modes, has been introduced as an *ad hoc* "fix" for this problem.[22]

Beyond these issues, both RPMD and CMD remain computationally relatively expensive due to the need to represent each physical atom in a simulation with dozens or more ring particle replicas (beads). Employing these methods on systems containing many thousands of atoms or more can be infeasible for all but the shortest simulations. For CMD, it has been found to be possible in some cases to generate accurate CMD dynamics with only a partial adiabatic separation of the internal ring polymer modes from the centroid mode.[23] Partially adiabatic CMD (PACMD) simulations can handle larger timesteps as well, allowing them to reach similar levels of efficiency to similar to RPMD. In the case of RPMD, however, a clever ring polymer contraction scheme can be used to increase computational efficiency.[24] In this method, the short range interactions of the system can be treated using a full ring polymer, while longer range interactions can be approximated by interactions calculated across a smaller number of beads (somethings even just the centroid mode). In order to further bridge the gap between classical MD and PIMD-based methods, approaches have been proposed[25,26] to directly evolve the centroids of quantum particles along a learned centroid force field, but these methods have used pairwise tabulated potentials in their effective quantum force fields which can limit their accuracy in capturing the NQEs.

To quickly summarize, CMD is by now a venerable approach to estimating certain quantum effects in finite temperature systems, for better or for worse. After nearly 30 years it has largely stood the test of time as a valuable approximation. Yet, the demanding nature of the calculation of the effective centroid force (usually done "on the fly") has in some ways held back the method from wider use and applicability (e.g., in comparison to RPMD). To shed light on a path to overcome this central challenge – and to capitalize on the rapidly evolving developments in machine learning (ML) – in this paper we introduce Machine-Learned CMD (ML-CMD) and demonstrate its (arguably remarkable) features in increasing the computational speed and overall efficiency (time to solution) for CMD simulations. The method employs a deep neural network (DNN) trained on PIMD data to act as a force field which calculates the effective centroid forces based on configurations of the path centroids alone. This method retains most of the efficiency of a pairwise force field without its corresponding constraints. ML-CMD employs the DeepMD kit,[27] which has been applied to both *ab initio* data as well as classical atomistic MD in the past to efficiently predict forces and energies of complicated systems.

DeepMD is a general use DNN method for learning atomic forces and energies based on two main components. The first is a descriptor network which converts the local environment, analogous to a neighbor list in classical MD, of a particle into translationally, rotationally, and permutationally invariant embeddings. This network then passes these embedded features to a second fitting network which considers this environment to predict atomic contributions to energy or force.[28,29] The method can learn a force field for a completely generic representation of a system using mapped forces, including those from PIMD trajectories.[30] In the case of ML-CMD, the training dataset consists of a PIMD trajectory with forces projected to the centroids of the imaginary time paths using the mean square error in the forces as a loss function. The dataset is then used to learn the centroid forces directly, thus front-loading the work of deriving the effective centroid forces. Naturally, this approach benefits simulations of larger systems the most, but even for simple applications it can result in significantly faster results. We show later in this paper that ML-CMD can be applied to low temperature para-hydrogen as well as room temperature liquid water in order to calculate quantum time correlation functions to a great degree of accuracy, as well as significantly greater efficiency (between one and two orders of magnitude faster) against comparable path integral based methods. The ML-CMD models are also easy to train, require no more additional simulation than CMD, and can be deployed quickly enough to justify their use over CMD and RPMD even for simple systems with no simulation data readily available.

**II. Methods**

We trained two ML-CMD models to test the method's ability to capture static and dynamic properties of systems with significant NQEs. The first is the Silvera-Goldman model[31] of para-hydrogen at 14 K and a density of $\rho=0.0235$ Å$^{-3}$. A total of 180 particles were simulated, each corresponding to an entire $H_2$ molecule. This spherical approximation is justified as hydrogen is in the rotational ground state at the chosen state point. The second system contained 233 water molecules simulated using the qSPC/fw force field at 300 K and atmospheric pressure.[32] The datasets used for DeepMD were generated using i-PI and LAMMPS software packages.[33,34] Each system was simulated using normal mode PIMD (NMPIMD) with 32 replicas per particle for 250 picoseconds. A total of 8000 frames were taken from the final 200 picoseconds of each simulation, and the coordinates and forces were mapped to the centroid resolution.

The para-hydrogen model was trained using an embedding DNN with three hidden layers containing 10, 20 and 40 neurons, no timestep, and 46 nearest neighbors were considered as the local environment for the descriptor network. The fitting network was composed of three hidden layers of 240 neurons each, with a timestep. The learning rate schedule was exponential, with a starting rate of $5\times10^{-3}$, and ending rate of $1.76\times10^{-7}$, and 5000 decay steps. The model was trained for 500,000 iterations before validation and testing. The water model was trained using an identical model except for the following differences: The first is the number of neighbors considered for the local environment of each atom was increased to 60 hydrogens and 30 oxygens. Second, the model was trained for 100,000 iterations instead of 500,000. Additionally, the water training was batched with a batch size of 10. Both models were trained using DeepMD's se_e2_a descriptor which uses pairwise distances and embeds both radial and angular information about the system into the network. In order to obtain the most efficient models possible, the training length and local environment size were systematically reduced until dynamical fidelity was impacted, thus resulting in a model which balances accuracy with speed. Input files for DeepMD training for both water and para-hydrogen are included in the Supporting Information.

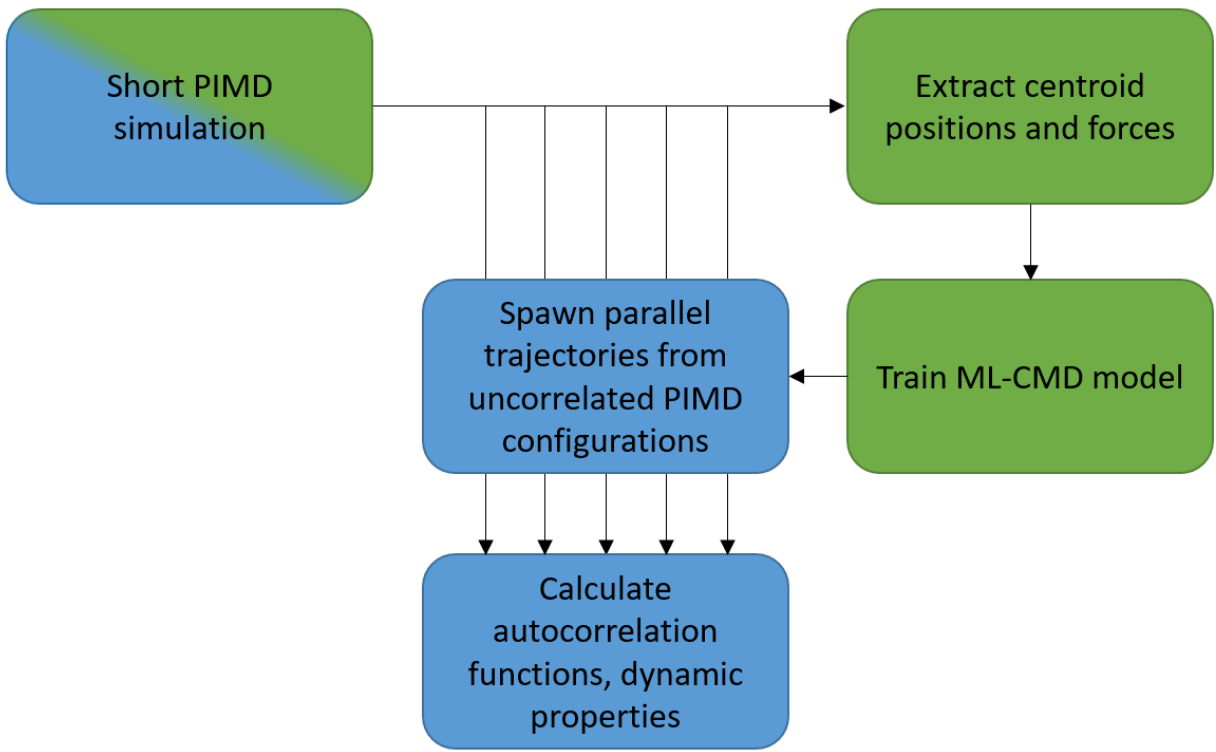

**Figure 1. Workflow for calculating autocorrelation functions using path integral MD methods and for generating ML-CMD models. Blue panels correspond to necessary steps for CMD and RPMD using uncorrelated configurations. Green panels are additional necessary steps for creating a ML-CMD model. The initial PIMD simulation step is shared among both, eliminating the cost of data collection for training ML-CMD force fields when calculating autocorrelation functions.**

The resulting force fields were tested against TRPMD and PACMD considering accuracy of static and dynamic properties as well as integration efficiency. All ML-CMD simulations were carried out in LAMMPS with the DeepMD force field add-on using a 0.5 fs timestep. For both RPMD and PACMD, each system was simulated using i-PI and LAMMPS using a 0.25 fs timestep for PACMD and a 0.5 fs timestep for TRPMD and 32 replicas per atom. PACMD frequencies were chosen according to the following equation

$$\Omega = P^{P(P-1)}/\beta\hbar, \tag{12}$$

which for the para-hydrogen system is 349 cm$^{-1}$ and for the water system is 7481 cm$^{-1}$.[21] In order to accurately compute diffusion and time correlation functions for these systems, a parallel scheme developed by Pérez et al. was used.[35] A total of 64 individual frames were selected from an initial PIMD simulation to generate uncorrelated starting points as shown in Figure 1. Each starting point was then used to launch two 6 ps TRPMD, PACMD or ML-CMD trajectories, for a total of 128 simulations for each method. Velocity autocorrelation functions (VACFs) were computed from the final 5 ps of each of these trajectories and averaged to obtain a converged result. Self-diffusion constants were then obtained from the zero-frequency Fourier transform of the Kubo transformed VACF. It is noteworthy that the PIMD trajectory used to start each short simulation can double as a dataset for ML-CMD. In cases where such a scheme is used, one saves even more time as this simulation performs two critical steps instead of just one.

When measuring efficiency, all methods were tested using the same computational environment. Each test simulation was run on 32 cores of a 40 core Cascade-Lake compute node to allow for one core per replica in the full PI simulations. All simulations were carried out in LAMMPS. The total time used to calculate autocorrelation functions was calculated as the sum of all required simulations and training ("time to solution"). For ML-CMD this includes the 250 ps PIMD simulation used to generate the dataset which was also used as the starting points for the autocorrelation function calculations, the DeepMD network training time, and the total time to run the 768 total ps of simulation required for the autocorrelation functions. For PACMD and TRPMD, this includes the short PIMD trajectory required to generate starting configurations and the time to simulate the 768 ps of data for the autocorrelation functions.

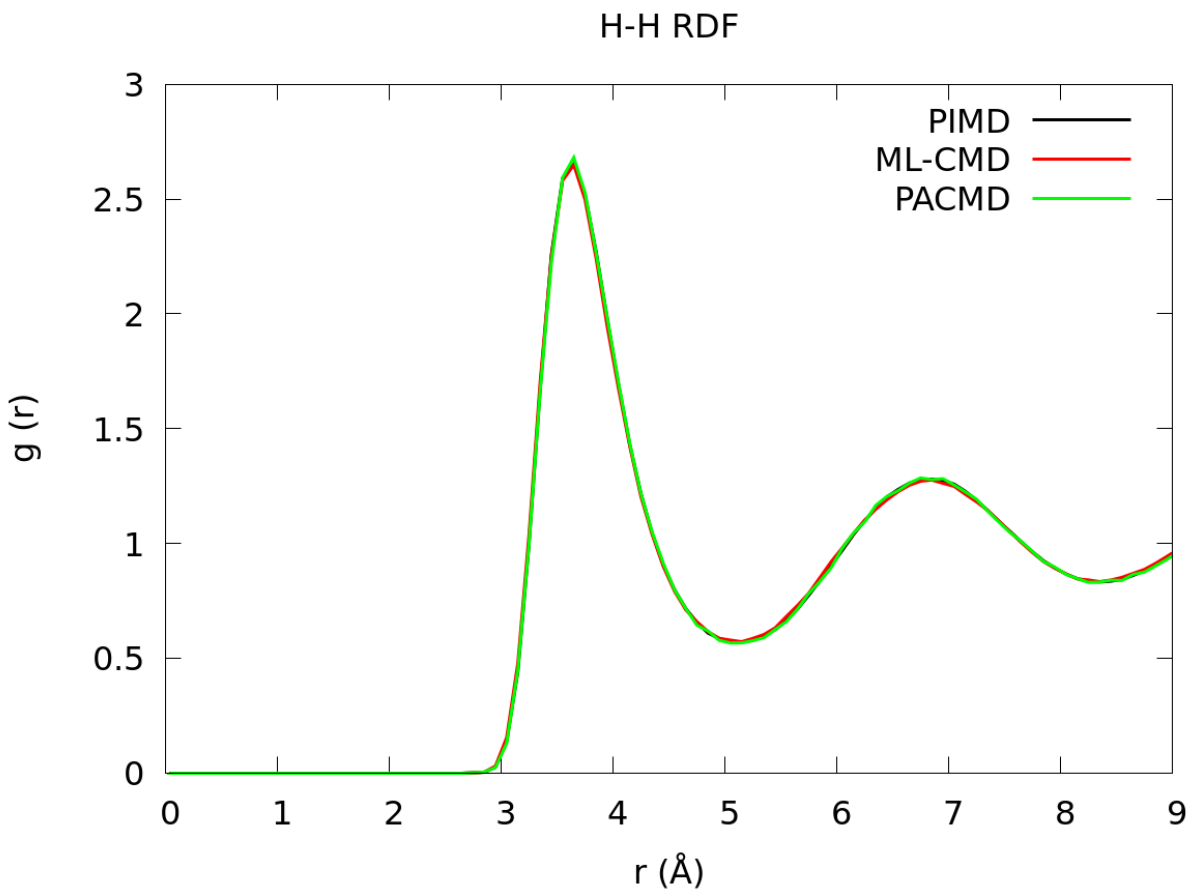

**Figure 2. Plot of the radial distribution function (g(r)) for 14 K para-hydrogen. PIMD, ML-CMD, and PACMD are compared. As the Silvera-Goldman potential represents one hydrogen molecule as a single particle no peak corresponding to the H-H bond is present.**

**III. Results and Discussion**

Both ML-CMD para-hydrogen and water results show excellent agreement with TRPMD and PACMD. Figures 2 and 3 show radial distribution functions for all three models as well as PIMD. Of note are the water peaks corresponding to the O-H bond and H-O-H angles, which ML-

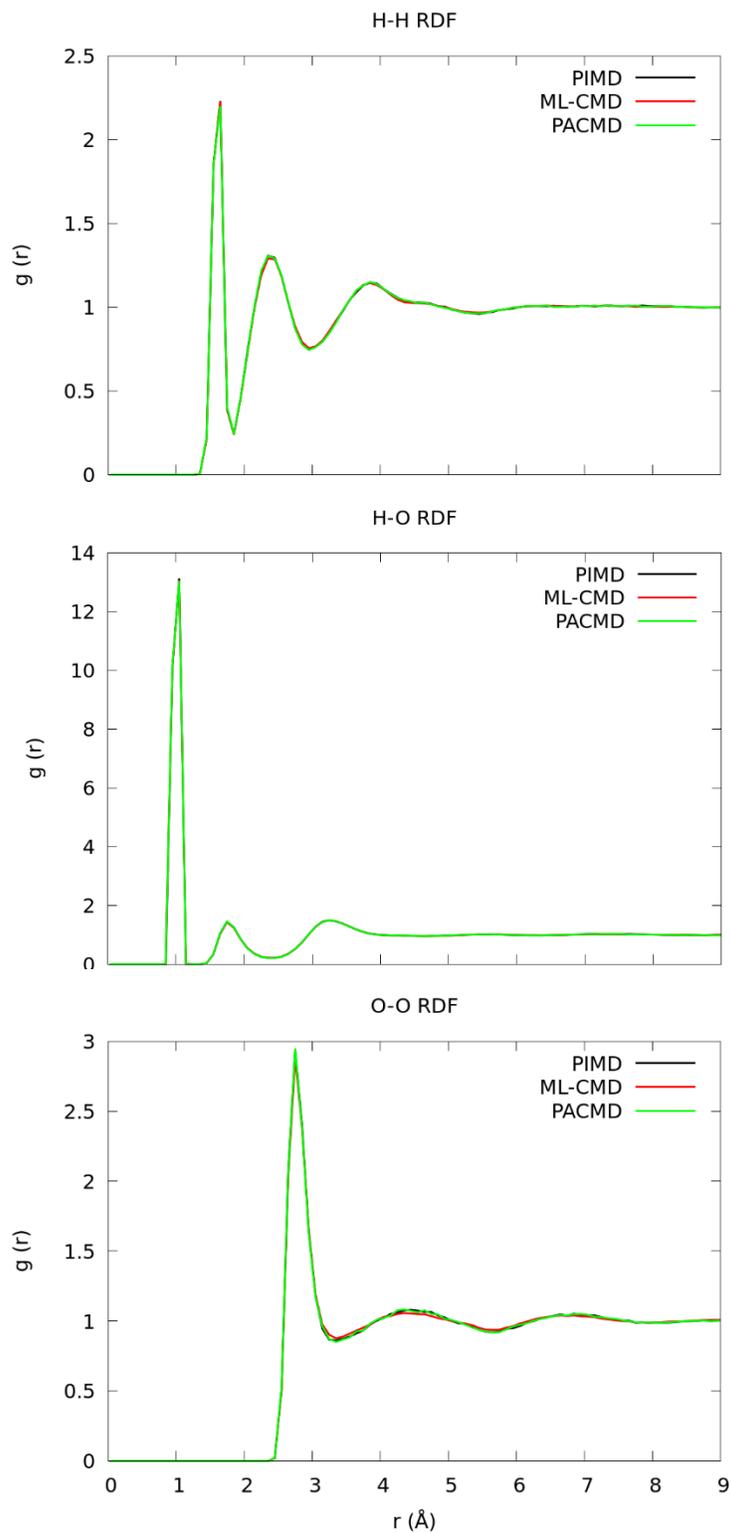

**Figure 3. Plot of the radial distribution function (g(r)) for Hydrogen-Hydrogen (top), Hydrogen-Oxygen (middle) and Oxygen-Oxygen (bottom) of 300 K water. PIMD, ML-CMD, PACMD, and TRPMD are compared.**

CMD captures very accurately despite there being no explicit bonded or angular interactions in the force field. It is, however, not surprising that the static properties of ML-CMD align with full PI methods as PIMD is well known to be ideal for sampling the equilibrium properties positions of quantum particles. In addition, a previous attempt at calculating effective centroid forces from PIMD trajectories for CMD have shown similar levels of accuracy for para-hydrogen.[25]

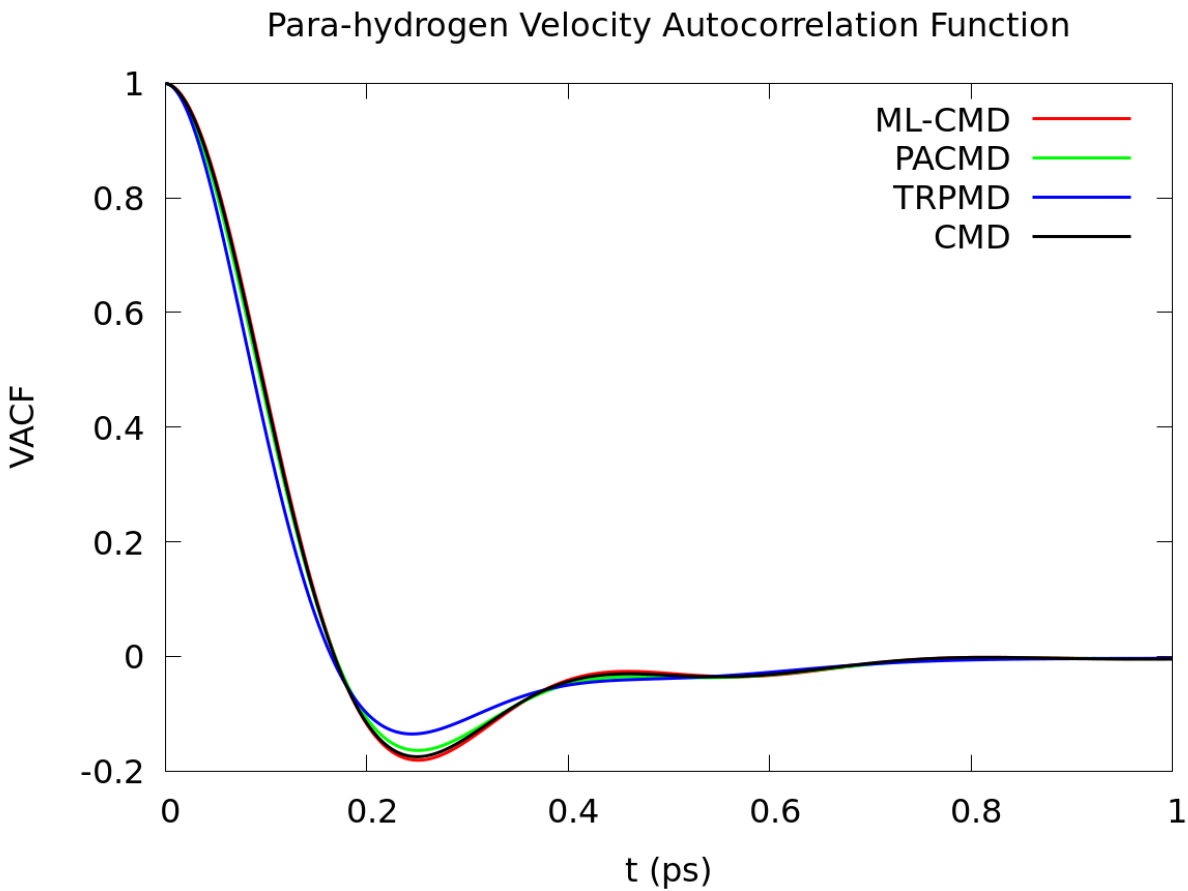

**Figure 4. A plot of the normalized velocity autocorrelation functions (VACFs) for 14 K para-hydrogen using ML-CMD, PACMD, CMD, and TRPMD simulations. Each VACF was averaged over 128 individual trajectories.**

The approximate quantum dynamics of the ML-CMD models are also well in line with the PACMD and TRPMD results. Figure 4 shows the VACF for all three types of simulation for para-hydrogen as well as CMD simulations with greater adiabatic separation. ML-CMD generally agrees better with PACMD than TRPMD, which is expected as the latter method is not designed to approximate centroid dynamics. CMD simulations in which the adiabatic separation is not partial show even better agreement with ML-CMD. This suggests that for systems with significant NQEs ML-CMD is not only faster than PA-CMD, but more accurate as well.

Figure 5 shows the velocity autocorrelation functions of water. In this case, ML-CMD matches PACMD and TRPMD nearly perfectly. The fact that RPMD and PACMD converge better for room temperature water than for 14 K para-hydrogen is not surprising. RPMD and CMD are known to converge as the system approaches the classical limit, e.g., heavier nuclei or higher temperatures.[23]

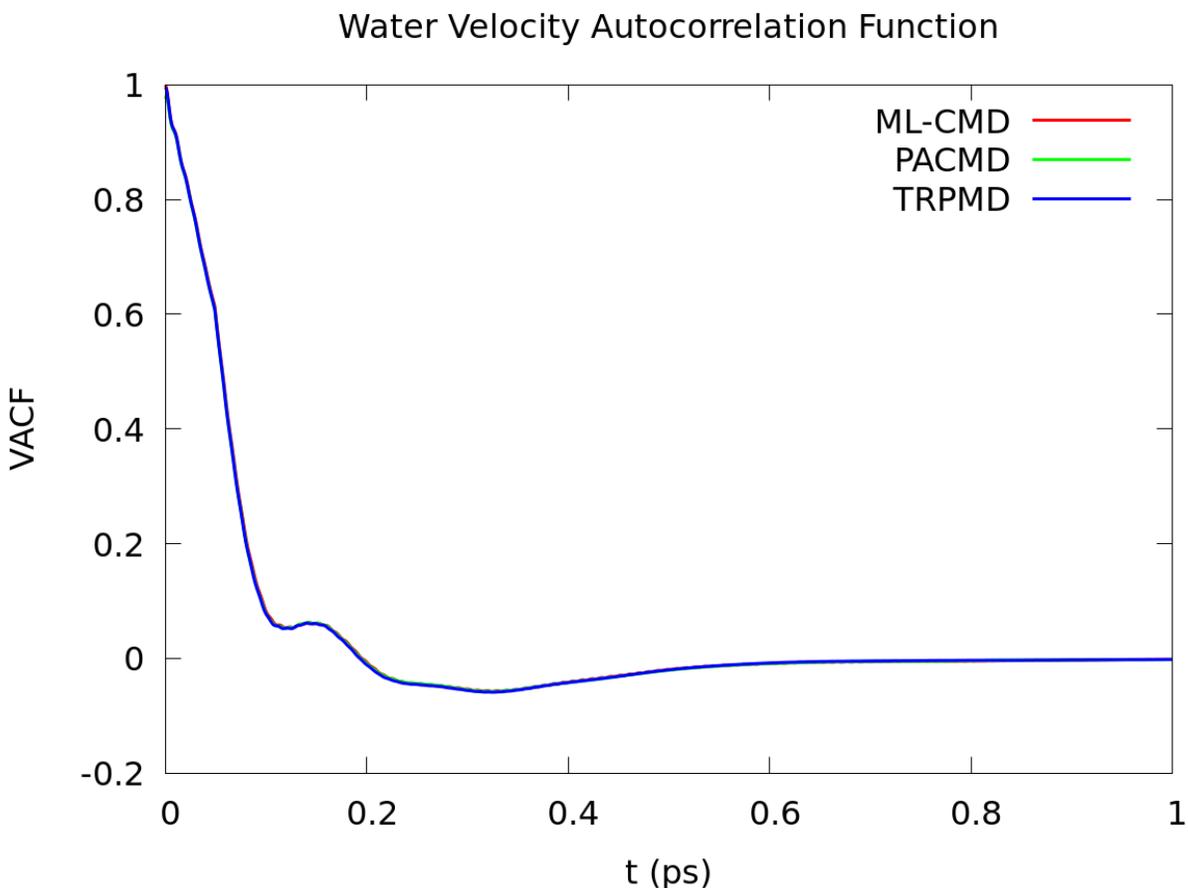

**Figure 5. A plot of the normalized velocity autocorrelation functions (VACFs) for 300 K using ML-CMD, PACMD and TRPMD simulations. Each VACF was averaged over 128 individual trajectories.**

Table 1 next shows the calculated self-diffusion constants for both water and para-hydrogen for all three types of simulation. ML-CMD shows excellent agreement in both cases; it appears to be slightly more diffusive although this is well within the margin of error. The neural network architecture of ML-CMD's force field seems to directly contribute to the dynamical accuracy of the method. A previous attempt for accelerating CMD using force matching of the centroid forces resulted in less accurate values for self-diffusion at lower temperatures.[25]

**Table 1. Self-diffusion constants for para-hydrogen and liquid water. All values were calculated using the zero frequency Fourier transform of the Kubo transformed velocity autocorrelation function.**

| System | Para-hydrogen 14 K | Water 300 K |
|---|---|---|
| ML-CMD | 0.30 ± 0.03 Å$^2$/ps | 0.32 ± 0.03 Å$^2$/ps |
| PACMD | 0.29 ± 0.03 Å$^2$/ps | 0.31 ± 0.03 Å$^2$/ps |
| TRPMD | 0.28 ± 0.03 Å$^2$/ps | 0.31 ± 0.04 Å$^2$/ps |

An important factor in the training and validation of DNNs is the amount of data required to achieve a converged model. The ratio of training data to extrapolation informs not only the feasibility of the method, but also its quality. We trained 3 additional para-hydrogen models using 20, 60, and 100 ps of PIMD reference data to test this. Table 2 shows diffusion coefficients for each of these models. Figure 6 shows VACFs comparing the models to the model trained on the full 200 ps reference dataset.

**Table 2. Diffusion coefficients for ML-CMD para-hydrogen trained on datasets of varying length compared to the base model, which was trained on 200 ps of PIMD simulation.**

| Reference simulation length | Diffusion Coefficient |
|---|---|
| 20 ps | 0.31 ± 0.03 Å$^2$/ps |
| 60 ps | 0.30 ± 0.03 Å$^2$/ps |
| 100 ps | 0.31 ± 0.03 Å$^2$/ps |
| 200 ps | 0.30 ± 0.03 Å$^2$/ps |

These results show that the method works with much less data than the full 200 ps trajectory. The diffusion constant values are well within a standard error of each other, and the autocorrelation functions match perfectly, even with 10 percent of the total training data used. Future applications of ML-CMD are thus likely to be deployed even more quickly and efficiently than the ones presented here.

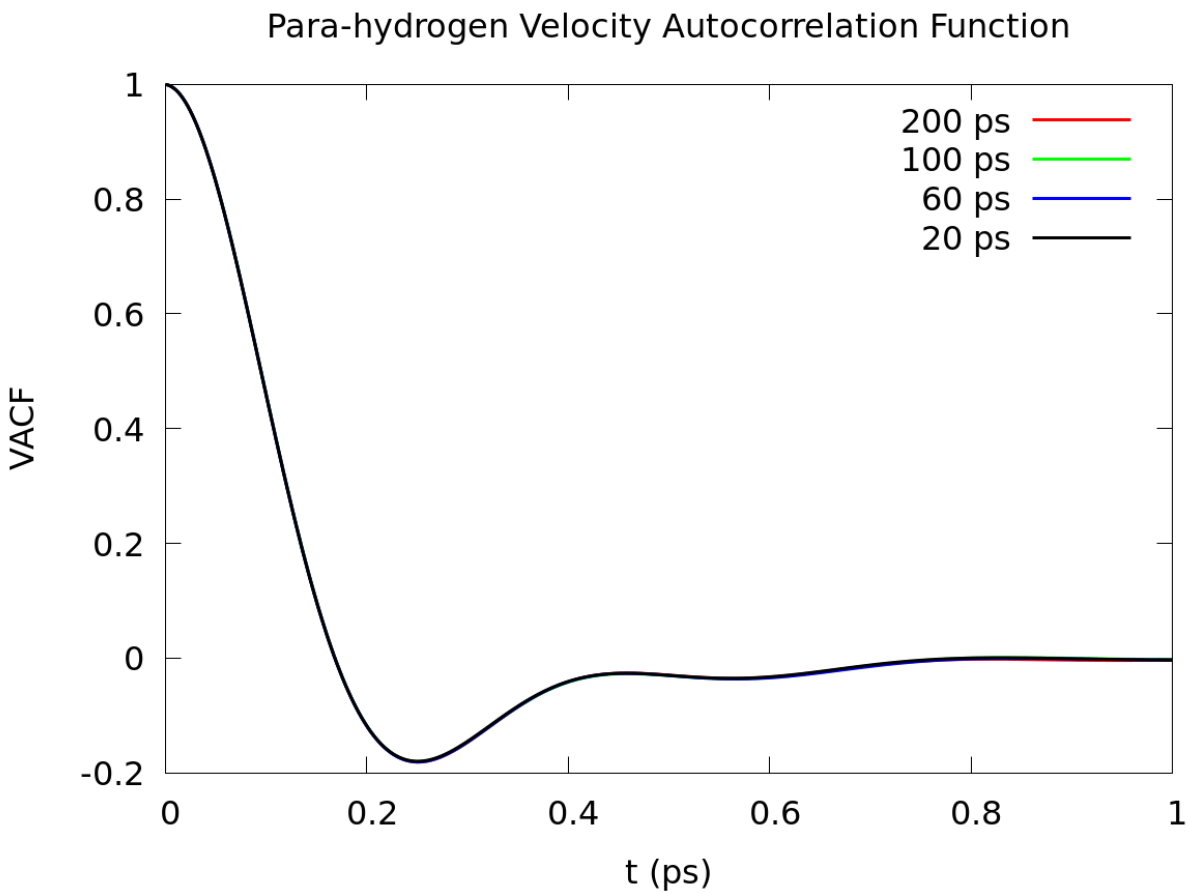

**Figure 6. A plot of the normalized velocity autocorrelation functions (VACFs) for 14 K para-hydrogen using several trained ML-CMD models. Each model was trained with a different amount of reference PIMD data and VACFs were calculated in the same manner as the preceding figures.**

Beyond accurate dynamics, the most important aspect of ML-CMD is its speed over methods such as full CMD, PACMD, RPMD, and TRPMD. In order to be useful as a replacement for these methods, it must be faster than them even when considering the time to collect the PIMD data and train the DNN force field. Table 3 shows integration speeds for the ML-CMD, PACMD, and TRPMD as well as the total time required to collect converged velocity autocorrelation functions. ML-CMD performs over 70 times faster than PACMD for para-hydrogen and over 20 times faster for water. Additionally, the time to data collection is shorter than both PI-based methods by a considerable amount, even when considering the amount of excess PIMD data used to train the models in the present case. These results are encouraging, as the correlation function calculations require only 768 total picoseconds of integration, less than four times the total amount of simulation time used to train the model in the first place. In cases where more or larger simulations are required such as calculating long time correlation functions for complex free energy surfaces, ML-CMD becomes even more efficient, as the initial simulations and model training only need to be performed once. Furthermore, each ML-CMD model can be shared for use in multiple simulations either individually or through a publicly available repository of models which can be contributed by anyone using the method.

Table 3. Integration speed and velocity autocorrelation function (VACF) calculation times for ML-CMD, PACMD and TRPMD. All measurements performed using 32 cascade-lake cores. VACF calculation time for PACMD and ML-CMD includes an initial PIMD simulation followed by 768 ps of integration. For ML-CMD, this also includes DeepMD neural network training time.

| System | Para-hydrogen speed | Para-hydrogen VACF calculation time | Water speed | Water VACF calculation time |
| --- | --- | --- | --- | --- |
| ML-CMD | 734.4 ps/hour | 21 hours | 144.5 ps/hour | 49 hours |
| PACMD | 9.8 ps/hour | 90 hours | 6.1 ps/hour | 137 hours |
| TRPMD | 17.9 ps/hour | 55 hours | 10.7 ps/hour | 84 hours |

It is also worth discussing the flexibility of the machine learning algorithm chosen for ML-CMD. DeepMD is simple to deploy and yields excellent performance, but it is one of many similar neural network based methods which can predict forces and energies in molecular systems. One of the most important aspects of any of these methods is how they encode translational, rotational and permutational invariances into the input features of the model. This allows for the network to work in scalar space with the same symmetries present in the real chemical environment. Recently, several models have been developed which replace these invariant networks with equivariant ones which can directly encode vectored information such as forces. These models have been shown to require far less data than symmetry invariant models, and can produce better results with up to one thousandth the number of training examples.[36,37] Applying such a method would naturally speed up the training of ML-CMD models and should be considered for future study.

**IV. Conclusions**

In this work, we have presented ML-CMD, a machine learning approach for the calculation of the effective centroid potential in CMD simulations. Over the past 30 years, regular CMD has been limited by the need to repeatedly calculate the centroid effective force "on-the-fly" through adiabatic separation of the centroid and non-centroid imaginary time path integral modes. Such calculations represent a significant computational overhead for larger systems. Instead, in the present work we shown that by training a neural network to first learn the centroid effective potential from a PIMD simulation, one can greatly increase the efficiency and time to solution of a CMD simulation without sacrificing accuracy.

We have demonstrated that for both room temperature water and 14 K para-hydrogen ML-CMD provides highly accurate results which closely match both the static and dynamic properties given by full PACMD and TRPMD simulations. The ML-CMD simulations are also many times faster than either of the latter approaches, thereby extending the range of systems for which CMD will be applicable. While the initial PIMD simulations and DNN training steps somewhat constrain the overall speed at which the ML-CMD models can be deployed, the PIMD simulations are also necessary to provide the initial conditions for trajectories in both PACMD and TRPMD.

This paper provides yet another example where machine learning promises to transform the field of molecular simulation, in this case by making CMD simulations feasible and accurate for a wider range of systems. Likewise, one can expect that future advances in machine learning, e.g., to better treat heterogenous systems, rare events, etc, will also provide clear benefit to the ML-CMD approach developed in this work.

## ASSOCIATED CONTENT

**Supporting Information**

The files used to train the para-hydrogen and water DeepMD models are included for use in replicating the data from this manuscript, as well as for training additional ML-CMD models.

## AUTHOR INFORMATION

**Corresponding Author**

**Gregory A. Voth**

Department of Chemistry, Chicago Center for Theoretical Chemistry, Institute for Biophysical Dynamics, and James Franck Institute, The University of Chicago, Chicago, IL 60637, USA

Email gavoth@uchicago.edu

**ORCID:** Gregory A. Voth 0000-0002-3267-6748

**Authors**

**Timothy D. Loose**

Department of Chemistry, Chicago Center for Theoretical Chemistry, Institute for Biophysical Dynamics, and James Franck Institute, The University of Chicago, Chicago, IL 60637, USA

**Patrick G. Sahrmann**

Department of Chemistry, Chicago Center for Theoretical Chemistry, Institute for Biophysical Dynamics, and James Franck Institute, The University of Chicago, Chicago, IL 60637, USA

**Acknowledgements**

This material is based on work supported by the National Science Foundation (NSF Grant CHE-2102677). Simulations were performed using computing resources provided by the University of Chicago Research Computing Center (RCC). T.D.L. was also supported by the National Science Foundation Graduate Research Fellowship (DGE-1746045). We thank Dr. Chenghan Li and Dr. Won Hee Ryu for helpful conversations on path integral simulations.

**For Table of Contents Only**

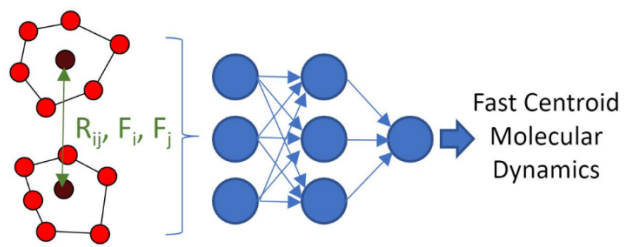